\documentclass[pra,aps,groupedaddress,twocolumn,floatfix,nofootinbib]{revtex4}
\usepackage{graphicx}

\newcommand{\beq}{\begin{equation}}
\newcommand{\eeq}{\end{equation}}
\newcommand{\beqa}{\begin{eqnarray}}
\newcommand{\eeqa}{\end{eqnarray}}

\newcommand{\braket}[2]{\mbox{$ \langle #1 | #2 \rangle $}}

\newcommand{\ket}[1]{\mbox{$ | #1 \rangle $}}
\newcommand{\bra}[1]{\mbox{$ \langle #1 | $}}

\def\opone{\leavevmode\hbox{\small1\normalsize\kern-.33em1}}

\begin{document}

\title{Collapse. What else?}
\author{Nicolas Gisin}

\date{\small \today}
\begin{abstract}
We present the quantum measurement problem as a serious physics problem. Serious because without a resolution, quantum theory is not complete, as it does not tell how one should - in principle - perform measurements. It is physical in the sense that the solution will bring new physics, i.e. new testable predictions; hence it is not merely a matter of interpretation of a frozen formalism. I argue that the two popular ways around the measurement problem, many-worlds and Bohmian-like mechanics, do, de facto, introduce effective collapses when ``I'' interact with the quantum system. Hence, surprisingly, in many-worlds and Bohmian mechanics, the ``I'' plays a more active role than in alternative models, like e.g. collapse models. Finally, I argue that either there are several kinds of stuffs out there, i.e. physical dualism, some stuff that respects the superposition principle and some that doesn't, or there are special configurations of atoms and photons for which the superposition principle breaks down. Or, and this I argue is the most promising, the dynamics has to be modified, i.e. in the form of a stochastic Schr\"odinger equation.
\end{abstract}
\maketitle

\section{The Quantum Measurement Problem}\label{QMP}
Quantum theory is undoubtedly an extraordinarily successful physics theory. It is also incredibly fascinating: somehow, by ''brute mental force'', one can understand the strange and marvelous world of atoms and photons! Furthermore, it is amazingly consistent, in the sense that it is amazingly difficult to modify the formalism: apparently, any change here or there activates non-locality, i.e. allows one to exploit quantum entanglement for arbitrary fast communication \cite{Gisin89,Gisin90,Gisin95}. How could the fathers  develop such a consistent theory based on the very sparse experimental evidence they had? In the landscape of theories, whatever that means, quantum theory must be quite isolated, so that if one looks for a theory in the neighborhood, one has to meet it. However, the quantum formalism is not consistent if one demands that a physics theory tells how one should, in principle, make measurements, as we develop in this article; it is also not consistent if one treats the observer as a quantum system \cite{RennerBohm}.

Quantum theory is a physics theory and all physics theory should tell what is measurable and how to perform measurements. About the first of these two points, quantum theory tells that all self-adjoint operators correspond to a measurable quantity. More precisely and probably more correct (it depends on the textbooks), quantum theory claims that every physical quantity is represented by a self-adjoint operator and every physical quantity can be measured (almost by definition of a physical quantity). Note that often these measurable physical quantities are called observables. So far so good. But let's turn to the second point above: how to perform measurements. Here quantum theory is surprisingly silent. Often it is said that one should couple the system under investigation to a measurement apparatus, frequently called a pointer, and then measure the latter \cite{VonNeumann32} (for a recent application of this, related to the measurement problem, see \cite{MQM16}). Hence, to measure a physical quantity of interest of your quantum system, you should measure another system. This is the infamous shifty split, as the pointer itself should be measured by coupling it to yet another measurement system, and so on.

If one insists, the theory remains silent. But the defender of the theory get virulent: ``If you don't know how to perform measurements'', they claim, ``then you are not a good physicist!''. Ok, physicists do know how to perform measurements, indeed, especially experimental physicists. But shouldn't all physics theories tell how to perform measurements, at least in principle?

Somehow, quantum theory is incomplete. I belong to the generation that learned that one should never write such a claim in a paper, at least if one wants to publish it in a respectable journal\footnote{or on the arxiv \cite{IQOQIblog}?}.
Admittedly, one has to be careful with such incompleteness claims. The idea is clearly not to go back to classical physics, i.e. to a mechanistic theory in which cogwheels and billiard balls push other cogwheels and billiard balls. The idea is also not to complement quantum theory with local elements of reality, using EPR's terminology \cite{Einstein35}, nor with local beables in Bell's terminology \cite{Bell87}. The idea is simply to complement quantum theory is such a way that it tells how, in principle, one performs measurements.

To illustrate the kind of complements I am looking for, one could, for example, postulate that the world is made out of two sorts of stuff, one to which the quantum mechanical superposition principle\footnote{The superposition principle states, in words, that if some stuff can be either in one state {\bf or} in another, it can also be in the first state {\bf and} in the second one, i.e. in superposition of the two states.} applies and one to which it doesn't apply. Quantum theory would describe only the first kind of stuff and measurements happen when one couples (somehow) the two sorts of stuffs. The readers will have recognized standard Copenhagen interpretation of quantum theory: the theory applies only to the ``small stuff'', while the superposition principle doesn't apply to the ``large stuff''. Measurements happen when one couples a small quantum system to a large measurement apparatus. As sketched here, this dualistic idea, i.e. that there are two kinds of stuffs, is not yet a complete theory. First, because it doesn't tell how to recognize the two sorts of stuffs (besides that the superposition principle applies to one and not to the other). Second, because it doesn't tell how to couple these two kinds of stuffs. Moreover, one may argue that a complete theory should also describe the other - non quantum - kind of stuff. Nevertheless, I believe that this line of thought deserves to be investigated more in depth, see section \ref{Dualism}.

What are the alternatives to some sort of dualism? Assume there is only one sort of stuff, but certain arrangements of this stuff make it special. For example, assume everything is made out of elementary particles, but certain arrangements of atoms and photons make them act as measuring apparatuses\footnote{Note that this would imply that the property of acting like a measurement apparatus is an emergent property. Moreover, once this emergent property obtains, the proper arrangement of atoms would gain the capacity of top-down causality, as such arrangements of atoms would have the power to collapse superpositions.}. Hence the question:\\ \\
{\it Which configurations of atoms and photons characterize measurement setups?}\\

Looking for such special configurations is an interesting line of research. Thanks to the world-wide development in quantum technologies, we should soon be able to investigate highly complex configurations of (natural or artificial) atoms and (optical or micro-wave) photons. Shall these developments lead to a breakthrough in quantum physics? Possibly, though most physicists bet on the contrary, i.e. bet that arbitrarily complex quantum processors will be developed, showing no sign of ``collapses'', no sign of any breakdown of the superposition principle.

\section{What is Physics}
Quantum theory explains very well why it is more and more difficult to keep coherence when the complexity increases: because of the so-called decoherence phenomena. So, are we facing the end of ``clean physics''? Shall we have to stay with the fact that, apparently, quantum theory holds at all scales, i.e. the superposition principle is truly universal, but for all practical purposes (FAPP, as Bell would have said \cite{BellFAPP}) there is a sort of complexity law of Nature - a sort of 2nd law -  that states that it is ultimately harder and harder to demonstrate the superposition principle experimentally for larger and more complex systems? Who knows. But for sure, we - the physics community - should not give up the grand enterprise that easily\footnote{In \cite{BellFAPP} John S. Bell wrote: ``In the beginning natural philosophers tried to understand the world around them. Trying to do that they hit upon the great idea of contriving artificially simple situations in which the number of factors involved is reduced to a minimum. Divide and conquer. Experimental science was born. But experiment is a tool. The aim remain: to understand the world. To restrict quantum mechanics to be exclusively about piddling laboratory operations is to betray the great enterprise. A serious formulation will not exclude the big world outside the laboratory.''\label{BellPhysics}}. Recall that \\

{\bf Physics is all about extracting information about How Nature Does It.}\\

And for physicists “extracting information” means performing measurements. Hence the measurement problem has to be taken seriously.

Let me stress that I consider the quantum measurement problem as a serious and real physics problem. It is serious because without a solution quantum theory is incomplete, as discussed above. It is real in the sense that it's solution will provide new physics, with new and testable predictions. Hence it is not merely a matter of interpretation of a given formalism: to solve it, one has to go beyond today's physics.

To conclude this section and to be transparent, I should state that I am a naive realist (as most physicists): there is a world out there and the grand enterprise of Physics aims at understanding it, see footnote \ref{BellPhysics}. Additionally, following Schr\"odinger \cite{SchrodingerDualism}, I consider that ``I'' am not part of it: my aim is to understand the outside world, but I am not including myself in that outside world. Of course, I am made out of atoms and other stuff that can and should be studied by physics. But physics is not about explaining my presence. As much as possible (and I believe it is entirely possible), physics theories should not postulate that ``I'' have to exist for the world to function. This may seem too philosophical, but we shall see that it has consequences for possible solutions to the measurement problem. Let me stress that this is not dualism in a physics sense: the world out there could well be made out of a single kind of stuff. 

In summary, I believe that the scientific method will never explain why there is something rather than nothing, nor will it explain why ``I'' am here. Physics must assume both that ``I'' exist and that there is a world out there, so that ``I'' can gain better and better understanding of the outside world, i.e. of How Nature Does It.


Let's return to the quantum measurement problem and look for alternatives to what we already discussed, i.e. physical dualism (the assumption that there are more than one sort of stuffs out there), to the existence of ``special'' configurations of atoms and photons that make them act like measurement devices, and to the end of clean physics\footnote{In order to avoid receiving a km-long e-mail from Chris Fuchs, let me say a few words about QBism \cite{qbism}. QBism changes the goal of physics. It is no longer about finding out How Nature Does It, QBism restricts physics to what ``I'' can say about the future. More precisely, about how ``I'' should bet on future events. For me this is not only a betray of the great enterprise, it is almost a sort of solipsism where everything is about ``me'' and my believes. Well, at the end, I am not sure I'll avoid the km-long e-mail.}.

\section{Many-Worlds}\label{ManyWorlds}
Why not simply assume that quantum theory is complete and the superposition principle universal? This leads straight to some many-worlds interpretations of quantum theory \cite{DeutschFabricReality97,Kent09}. Indeed, since quantum theory is amazingly successful and since quantum theory without any addition (i.e. without any vague collapse postulate) leads to the many-worlds, why not merely adopt a many-worlds view? 

In the many-worlds view, the measurement problem is circumvented by the claim that everything that has a chance to happen, whatever tiny chance, does actually happen. Hence, it is a sort of huge catalog of everything that could happen. More precisely, it is the catalog of everything that has happened and of everything that is happening and of everything that will ever happen. Simply, we are not aware of the entire catalog, only of that part of the catalog corresponding to the world in which ``we'' happen to live in. But isn't physics precisely about, and only about, that part of the catalog? What is the explanatory power of claiming that everything happens, but ``we'' are not aware of everything? And what is that ``we''? 

It is a fact that ``I'' exist. Actually, it is the fact  that I know best. Should ``I'' be satisfied with a theory that tells that I exist in a hugely enormous number of copies and that all the theory provides is a catalog of everything that ``I'' or a copy of myself experiences? Not to mention the vast majority of worlds in which the atoms of my body don't make up a human, probably not even a thing. Actually, the theory says a bit more, it also tells about correlations. If ``I'' see this now, then there are only some events ``I'' may see in the future. And vice-versa, as time doesn't properly exist in the many-worlds. Note that to achieve this, one conditions the catalog on what ``I see now'', i.e. one uses an effective collapse: one limits the analysis to that part of the catalog in which ``I'' see this now. 

In summary, in many-worlds theories, it is ``I'' that continuously collapse the state-vector, at least for the purpose of allowing the theory to make predictions about what ``I'' am observing. In other worlds, in many-worlds the ``I'' is not merely a passive observer, but plays an active role.

Admittedly, many-worlds is a logically consistent interpretation, at least as long as one doesn't insist that ``I'' exists. Moreover, it is the most natural one if one sticks to standard Hilbert-space quantum theory (i.e. without measurements). But logical consistency is only a necessary condition for a physics theory. Solipsism is another example of a logically consistent theory, somehow on the other extreme to many-worlds: in solipsism only ``I'' exists. But, as I stated in the previous section, I am a realist: I just don't see how one can do physics without assuming the ``I'' and the ``world''.

Let me address another issue with many-worlds. It is a deterministic theory, even a hyper-deterministic theory, i.e. determinism applies to everything in the entire universe. Indeed, since there can't be any influences coming from outside and since the Schr\"odinger equation - the only dynamical equation of the theory - is deterministic, everything that happens today, e.g. what I am writing, the way each reader reacts, the details of all solar eruptions, etc, was all encoded in some ``quantum fluctuations''\footnote{Don't ask me what that means.} of the initial state of the universe\footnote{or, equivalently, in the final state of the Universe.}. Given the complexity of the (many-) worlds, it had to be encoded in some infinitesimal digits of some quantum state, possibly in the billionths of billionths decimal place. I am always astonished that some people seriously believe in that. Mathematical real numbers are undoubtedly very useful when doing our theory. But are they physically real \cite{GisinTimePasses}\footnote{recall that the assumption that real numbers are physically real implies that there could be an infinite amounts of information in a finite volume of space \cite{Chaitin,GisinTimePasses}.}? Do these infinitesimal digits have a real impact on the real world? Is this still proper physics? For sure, such assumptions can't be tested. Hence, for me, hyper-determinism is a non-sense \cite{GisinTimePasses}, though it is the dominant trend in today's high-energy physics and cosmology (though see \cite{SmolinTimePasses,NortonTimePasses}). Apparently, the many followers of today's trend elevate (unconsciously) the linearity of the Schr\"odinger equation and the superposition principle to some sort of ultimate quasi-religious truth, some truth in which they believes even more than in their own free will. Note that it is not the first time in science history that some equations get elevated above reason: followers of Laplace did also elevate the deterministic Newton equations to some sort of ultimate truth. We know what was the destiny of that belief.

In summary, in order to make predictions in the many-worlds, one introduces some effective collapses that happen when the system is coupled to ``I''. Hence, the theory is not complete, but relies - somehow - on ``I'', i.e. on some concept foreign to the theory.

\section{Bohmian Quantum Mechanics}\label{Bohm}
There is yet another way to avoid the quantum measurement problem. Assume that at all times there is one and only one ``event'' that is singled-out\footnote{or one collection of events that are singled-out. We may name this collection as one ``big'' or ``composed'' event.}. As time passes, the list of singled-out events must be consistent, as in consistent histories \cite{DH}. A nice example assumes that, at the end of the day, everything we ever observe is the position of some stuff. Hence, let's assume that the physical quantity ``position'' is always well determined by some additional variable (additional with respect to standard quantum theory). Interestingly, this can be made consistent \cite{Bohm52,BohmHiley}, though at the cost of some counter-intuitive phenomena \cite{Englert92,GisinBohm15} and assuming it applies to the entire universe (as soon as one cuts out some piece of the universe, one may encounter paradoxes \cite{RennerBohm}). 

Note that one may also apply similar ideas to other physical quantities than position, leading to various modal interpretations of quantum theory \cite{VanFraassenModal80,modal}. With position as the special physical quantity, the reader has recognized Bohmian quantum mechanics \cite{Bohm52,BohmHiley}. It is a nice existence proof of non-local hidden variables that deserves to be more widely known \cite{BellBohm}. It is non-local despite the fact that the additional variables are points in space, i.e. highly localised. But the dynamics of these point-particles is non-local: be acting here one can instantaneously influence the trajectories of point-particles there, at a distance. At first, this might be considered as quite odd. But quantum physics is non-local, in the sense of violating Bell inequalities. Hence, the non-locality of Bohmian mechanics is quite acceptable. Actually, there is just no choice: in order to recover the predictions of quantum theory and the experimental data, all theories must incorporate the possibility of Bell inequality violations, i.e. some non-locality.

One ugly aspect, in my opinion, of Bohmian mechanics is that the additional variables must remain hidden for ever. If not, if one could somehow collect information about their locations beyond the statistical predictions of quantum theory, then one could activate non-locality, i.e. one could use entanglement not only to violate some Bell inequality, but to send classical information at an arbitrarily large speed \cite{Valentini91}. But can one add variables to a physics theory while claiming that they are ultimately not accessible? Bohmians answer that the hidden positions determine the results of measurements, hence are not entirely hidden. Indeed, when one observes a result, one can apply an effective collapse as one knows that the hidden positions are now distributed within the reduced wave-function corresponding to quantum statistics. But there is no way to know more about the location of that particle\footnote{Moreover, there are situations in which the hidden particle leaves a trace where it was not \cite{Englert92,GisinBohm15}.}. This raises the question when should one apply such an effective collapse? The answer presumably is: ``when ``I'' register a measurement result'', a bit like in many-worlds.

I find it tempting to compare Bohmian mechanics with a toy-theory in which one has added as additional variables all the results of all the measurements that will ever be performed in the future, though with the restriction that none of these additional variables can be accessed before the corresponding measurements take place. Note that in such a way one can turn any theory into a deterministic one\footnote{Note though, that with the sketched construction, time is necessarily build into the toy-theory.}. But, for sure, no physicists would take such a toy-theory seriously. Admittedly, Bohmian mechanics is much more elegant than the above sketched one. But is it fundamentally different? 

Let's return to Bohmian quantum mechanics. As said, it is a remarkable existence proof of non-local hidden variables. But does it answer the deep question of the quantum measurement problem? I don't think so. As with the many-worlds, it assumes hyper-determinism and relies on infinitesimal digits for its predictions. Hence, despite the deterministic equations, it is not a deterministic theory \cite{GisinTimePasses}, as I elaborate in the next section. Moreover and disappointingly, it doesn't make any new prediction.

Finally, Bohmian mechanics with its non-local hidden variable is at great tension with relativity.

\section{Newtonian Determinism}\label{Determinism}
Some readers may wonder whether I would also have argued against classical Newtonian mechanics, had I lived 150 years ago. After all, it also relies on deterministic equations and Newton's universal gravitation theory is also non-local; and what about the ``I''. Let me start with the second aspect, non-locality. I have no problem with quantum non-locality (the possibility to violate Bell inequalities), because quantum randomness precisely prevents the possibility to use quantum non-locality to send classical information \cite{PopescuRohrlich94,Gisin98, SimonBuzekGisin, BrunnerRMP14}. However, Newton's non-locality\footnote{which, by the way, predicts the possibility to violate Bell's inequality.} can be used, in principle, to send information without any physical support carrying this information: move a rock on the moon (with a small rocket) and measure the gravitational field on earth. According to Newton's theory this allows one to communicate in a non-physical way - i.e. without any physical stuff carrying the information - and at an arbitrarily high speed \cite{GisinQchance14}. This is deeply disturbing and did already disturb Newton himself \cite{Newton}:

{\it That Gravity should be innate, inherent and essential to Matter, so that one Body may act upon another at a Distance thro’ a Vacuum, without the mediation of any thing else, by and through which their Action and Force may be conveyed from one to another, is to me so great an Absurdity, that I believe no Man who has in philosophical Matters a competent Faculty of thinking, can ever fall into it.}

Admittedly, when I first learned about Newton's universal gravitation theory at high school I found it beautiful, not noticing how absurd it is. May I suggest that one should always teach Newton together with the comment that it is efficient but absurd? I believe this would be great pedagogy.

Today we know that relativity solved the issue of Newton's non-locality and that experiments have confirmed quantum non-locality beyond any reasonable doubts \cite{Hanson15,Zeilinger15,Nam15}. 

Let's now turn to the other similarity between Bohmian mechanics and classical physics, that is the deterministic nature of Newton's equations. For clocks, harmonic oscillators and generally integrable dynamical systems, the stability is such that the infinitesimal digits of the initial condition do play no role. For chaotic systems, on the contrary, these infinitesimal digits quickly dominate the dynamics. Hence, since these mathematical infinitesimal digit do not physically exist\footnote{Several colleagues complained that this "physical existence" is badly defined and/or confuses physical existence with measurability. Let me try to clarify. Obviously real numbers can't be measured, neither today nor in any future. But my claim goes way beyond that. The world out there is pretty well described by today's physics. Actually, it is also pretty well described by the physics of one or two centuries ago, and will be even better described by the physics in some centuries. But this doesn't allow us to identify the world out there with its physical description. The world out there is infinitely richer than any physical description and than any human description. Somehow, the world out there is "`free"', i.e. it doesn't let itself get trapped in our theories, it does not depend on our description. In particular the fact that we use real numbers doesn't imply that real numbers govern the world out there, nor does the fact that some of our descriptions are based on deterministic equations imply that the evolution of the world out there is deterministic. In brief, the world out there can't be confined (locked up) in any finite-time physics theory.}, chaotic dynamical system are not deterministic\footnote{One may argue that there is no quantum chaos. But this is not entirely true, though quantum chaos differs deeply from classical chaos. In the quantum case it is the high sensitivity to the exact Hamiltonian that should be considered. One may claim that there is one exact fundamental Hamiltonian, thus no indeterminacy. But this is wrong, since, whatever units one chooses, the Hamiltonian contains constants, like the masses of particles, and these constants are described by real numbers. These real numbers are themselves indetermined (random), hence even the fundamental Hamiltonian leads to chaos.}. This fact doesn't change anything in practice (FAPP, as Bell would have shouted \cite{BellFAPP}), but it demonstrates that classical Newtonian mechanics is simply not a deterministic physics theory: despite the use of deterministic equations, it does not describe deterministic physics \cite{GisinTimePasses}. 

There is, however, a huge difference between Newton's determinism (of the equations) and Bohmian or many-worlds. In the former there is no entanglement. Hence, one can separate the world into systems, hence ``I'' can act on each of them individually. In an enormously entangled world, on the contrary, there is no way to separate sub-systems, there is no way to act on just one sub-system. Determinism plus entanglement make things intractable \cite{GisinHasard10}. Accordingly, either ``I'' can not act, or ``I'' do induce effective collapses that disentangle the subsystems. But then, why not include these effective collapses in the theory?

\section{Dualism}\label{Dualism}
In summary, so far we saw 4 sorts of attempts to circumvent the quantum measurement problem:
\begin{enumerate}
\item dualism as in orthodox Copenhagen quantum mechanics, 
\item some configurations of atoms and photons make them act as measurement setups that break superpositions,
\item all possible results co-exist in some many-worlds, 
\item all results were already encode in some additional non-local variables, hidden for ever, as in Bohmian mechanics.
\end{enumerate}

Let me recall that since I consider the quantum measurement problem as a real physics problem, it's solution will necessarily lead to new physics, including new and testable predictions. It is a fact that so far attempts 2, 3 and 4 did not bring up as much good and new physics as attempt 1 did. But I should add that this argument might be a bit unfair, because attempt 1 came first and had thus a significant advantage. Anyway, let's consider attempt number 1, i.e. dualism.

It is probably fair to say that most physicists would reject dualism\footnote{By the way, many would even do so virulently, while at the same time claiming to adhere to the Copenhagen interpretation, which is dualist. The sames would simultaneously claim with joy how proud they are to work in a field where rational thinking dominates. Ok, I leave that line of thoughts to sociologists.}. But could it be that they go too fast here? Clearly, dividing the world out there into ``small'' and ``large'' is not good enough. But couldn't there be stuff to which the superposition principle doesn't apply? Some have argued that the hypothetical non-quantum stuff is space-time and/or gravity \cite{Karolyhazy90, Diosi87,Penrose96,Penrose03,Adler07}. This is certainly a possibility. But I am reluctant to put my bets on this, because everything is connected to space-time and to gravity. Hence, if it is the coupling between the ``quantum stuff'' and the hypothetical ``non-quantum stuff'' that determines when a measurement happens, then, continuously, everything always undergoes measurements. In such a case, either the superposition principle is continuously broken and one should never have seen superpositions, or the non-quantum stuff undergoes a bit of superposition. 

More formally, denoting $\ket{QS_0}$ the initial state of some quantum stuff that interacts with some non-quantum stuff $\ket{NQS_0}$, then, after an arbitrary short time the quantum and non-quantum stuff get entangled:
\beq\label{entStuff}
\ket{QS_0}\ket{NQS_0}\stackrel{t=\epsilon}{\rightarrow}\sum_j\ket{QS_j}\ket{NQS_j}
\eeq
But if the non-quantum stuff can't at all be in superposition, then state (\ref{entStuff}) can't exist, not even for a split of a second. Hence, there would be instantaneously collapse also for the quantum stuff.

People have speculated that this bit of superposition gets quickly, though not instantaneously, washed out \cite{Karolyhazy90, Diosi87,Penrose96,Penrose03,Adler07}. Why not. But then, why introduce such a non-quantum stuff in the first place? Why not merely assume that all stuff undergo superpositions, but only in some (precisely) limited way? Readers recognize here spontaneous collapse theories; more on this in section \ref{Collapse}.

Before closing this section, let's see whether there is not another plausible way to divide the stuff into several sorts, i.e. dualism\footnote{Physics divides ``me'' and the outside world. I do not consider this as fundamental dualism, but only as the scientific method. Here I am asking whether a real physical dualism is a viable path towards a resolution of the quantum measurement problem.}. There is obviously one that goes back all the way to Descartes: ``material stuff'' and ``non-material stuff''. The superposition principle would apply only to the material stuff. This is admittedly extremely crude, certainly not yet a theory, not even a valid sketch of a theory, because essentially nothing is said about the ``non-material stuff''. Moreover, one should not make the situation more confused by thinking that the ``non-material stuff'' is our ``mind'', as this would imply that the first measurement that ever happened had to wait for us. However, I like to argue that one should also not reject dualism too quickly. After all, it might well be that there is stuff out there to which the superposition principle does not apply. 


Let's return to the quantum measurement problem. Although I am sort of a dualist from a philosophical point of view\footnote{I don't believe that everything is merely matter and energy as described by today's physics, not even stuff described by any physics theory at any given point in time. Though I believe in endless progress.}, I don't think that dualism is the right solution for the measurement problem. It might be that in some decades, if the measurement problem remains without significant progress, one may have to revisit a dualistic solution, but at present we better stick to the assumption that there is one and only one sort of stuff out there in the real world and that the superposition principle applies to it.

\section{Modified Schr\"odinger Equation}\label{Collapse}
Recall that the superposition principle states, in words, that if some stuff can be either in one state {\bf or} in another, it can also be in the first state {\bf and} in the second one, i.e. in superposition of the two states. The linearity of the Schr\"odinger equation implies then that such superposition last for ever. Consequently, in a theory without the measurement problem and in which everything (except ``I'') satisfy the superposition principle (and without hyper-determinism) it must be the case that it is the Schr\"odinger equation that has to be modified. First attempts to modify the Schr\"odinger equation tried to extend it to some non-linear but still deterministic equation \cite{Kostin75,Birula76,GisinJPA81}. But this turned out to be hopeless, as could be expected from the discussions in the previous sections. A quite convincing argument came from the observation that any such deterministic nonlinear generalization of the Schrodinger equation activates non-locality, i.e. predicts the possibility of arbitrarily fast communication \cite{Gisin89,Gisin90,Gisin95}. 

Hence, one has to go for a non-deterministic generalization\footnote{Note that sticking to a linear equation is also hopeless, as the Schrodinger equation is the only linear equation that preserves the norm of the state vector, see also \cite{Wigner59,GisinJMP83}}. Non-deterministic merely means not deterministic, that is it does not say how the equation should be, it only says how the equation should not be. However, assuming that the evolution is Markovian and the solution continuous in time, then - for those who know stochastic differential equations - possibilities are quite easy to find \cite{Gisin84,Gisin89,DiosiPLA88,Pearle89}. Essentially there is only one \cite{PercivalBook}. This solution depends on some operators, a bit like the Schr\"odinger equation depends on the Hamiltonian. At this point, all that remains is to fix this operator of the new, non-linear and stochastic term of the hypothetically fundamental dynamical equation of the complemented quantum theory and look for the new predictions.

Let's be a bit more explicit. Consider the following It\^o stochastic differential equation, see e.g. \cite{Gisin84,GisinPercival}:
\beqa\label{dpsi}
\ket{d \psi_t}&=& -i H \ket{\psi_t} dt \nonumber\\
&+& \sum_j \left(2 \langle L_j^{\dag} \rangle_{\psi_t} L_j - L_j^{\dag} L_j - \langle L_j^{\dag} \rangle_{\psi_t} 
\langle L_j \rangle_{\psi_t} \right) \ket{\psi_t} dt \nonumber\\
&+& \sum_j \left( L_j - \langle L_j \rangle_{\psi_t} \right) \ket{\psi_t}d \xi_j
\eeqa
where $H$ is the usual Hamitonian, $L_j$'s are (Lindblad linear) operators, $\langle L_j \rangle_{\psi_t}=\frac{\bra{\psi_t}L_j\ket{\psi_t}}{\braket{\psi_t}{\psi_t}}$ are the expectation values of the operators $L_j$ and the $d\xi_j$'s are independent complex Wiener processes satisfying:
\beqa
 M [ d\xi_j ] &=& 0 \\
 M[ d \xi_j d \xi_k ] &=& 0 \\
 M [ d \xi_j d \xi_k^* ] &=& \delta_{jk} \ dt 
\eeqa
where $M[...]$ denotes the mean value. Note that eq. (\ref{dpsi}) preserves the norm of $\ket{\psi_t}$.

Equation (\ref{dpsi}) describes a sort of Brownian motion in Hilbert space of the state-vector $\ket{\psi_t}$. It is the analog of a stochastic description of Browian motion at the individual particle level. It is assumed that it is not merely an approximation, but the foundamental dynamical law describing how isolated quantum systems evolve. Hence it predicts deviations from the standard Schr\"odinger dynamics, i.e. it predicts new physics. Consequently, at least, such modified dynamical laws could be wrong! 

To illustrate eq. (\ref{dpsi}) and for simplicity, let's consider the case with a single operator $L$, furthermore assume it is self-adjoint and commutes with the Hamiltonian $H$. Then, interestingly, the solutions to (\ref{dpsi}) follow a sort of Brownian motion and eventually tends to an eigenstate $\ket{l}$ of $L$. Moreover, the probability to tend to a given $\ket{l}$ equals the quantum probability $|\braket{l}{\psi_0}|^2$, see Fig. 1. 
\begin{figure}
\includegraphics[width=9cm]{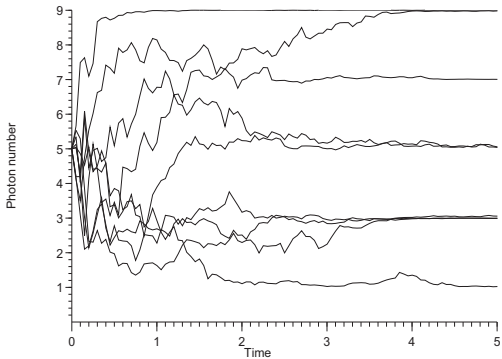}
\vspace{-30pt}
\caption{\it Example of some solutions to eq. (\ref{dpsi}) in case of a photon-number measurements, i.e. $H=L=a^\dagger a$. The initial state is an equal superposition of odd photon-number states: $\ket{1}+\ket{3}+\ket{5}+\ket{7}+\ket{9}$. The convergence to the eigenstates can be clearly seen. Taken from \cite{GisinPercival}.}
\end{figure}

When averaging over all solutions of (\ref{dpsi}), i.e. averaging over all Wiener processes $d\xi_j$, one obtains a density matrix $\rho(t)$ that satisfies the linear evolution equation:
\beqa\label{rhot}
{d \rho(t) \over dt} &=& -i  [H, \rho(t)] \\
&-& \sum_{j} \left(
L_j^{\dag} L_j\rho(t) + \rho(t)L_j^{\dag} L_j - 2 L_j \rho(t) L_j^{\dag} \right) \nonumber
\eeqa
Equation (\ref{rhot}) is the quantum analog of a classical Fokker-Planck equation describing the probability distribution of an ensemble of classical Browian particles.

Let us emphasize that since the density matrices at all times follow a closed form equation, this modification of the Schr\"odinger equation does not lead to the possibility of faster than light communication \cite{Gisin84,Gisin89}.

Remains to find what the operators $L_j$ could be. Here comes the beautiful finding of Ghirardi, Rimini and Weber \cite{GRW86}\footnote{In 1988 Professor Alberto Rimini visited Geneva to present a colloquium. He presented the famous GRW paper \cite{GRW86} in the version Bell gave of it \cite{BellGRW87}.
In the GRW theory, the non-linear stochastic terms added to the Schr\"odiger equation lead to solutions with discontinuous jumps of the wave-packet, i.e. to some sort of spontaneous collapses triggered by nothing but mere random chance, as time passes. Near the end of his colloquium, Rimini mentioned that an open question was to massage the stochastic modifications in such a way that the solutions would be continuous trajectories (in Hilbert space). He also emphasized the need for an equation that would preserve (anti-)symmetric states. He may have added that, with Philip Pearle \cite{CSL90}, they have a solution, but for sure he had no time to explain it. Immediately after the colloquium I went to Alberto and told him that I knew how to answer his questions. He encouraged me and I immediately added a small section to a paper already quasi-finished \cite{Gisin89}. There is no doubt that Philip Pearle found CSL independently. Lajos Diosi, by the way, did also find it \cite{DiosiPLA88}. Actually, everyone who, at that time, knew both GRW and It\^o stochastic differential calculus would have found it, because it is quite trivial, once you know the tools and the problem. Anyway, Ghirardi and Pearle got very angry that I published my result first and I decided to leave that field. I didn't like fights and wanted a carrier.}.
Assume the $L_j$ are proportional to the positions of all elementary particles, with a proportionality coefficient small enough that it barely affects the evolution of systems made out of one or only a few particles. Hence, microscopic systems would essentially not be affected by the modified Schr\"odinger equation (\ref{dpsi}). However, if a pointer is in superposition of pointing here and pointing there, then, since the pointer is made out of an enormous number of particles, let's say about $10^{20}$, the modified Schr\"odinger equation predicts a quasi instantaneous collapse: it suffices that a single particle gets localized by the stochastic nonlinear terms of equation (\ref{dpsi}) for the entire pointer to localize, i.e. the pointer localizes about $10^{20}$ times faster than individual particles. 

I remain convinced that collapse models of the form sketched above, see \cite{BassiRMP13} and references there in, is the best option we have today to solve the unacceptable quantum measurement problem. Note, however, the following two critical points.

First, one unpleasant characteristic of such a modified dynamics is that the very same equation (\ref{dpsi}) can also be derived within standard quantum theory by assuming some coupling between the quantum system and its environment and conditioning the system's state on some continuous measurement outcomes carried out on the environment \cite{Wiseman00}. This makes it highly non trivial to demonstrate an evolution satisfying equation (\ref{dpsi}) as a fundamental evolution, as one would have to convincingly show that the system does not interact significantly with its environment. Note also that sufficient error corrections could hide the additional stochastic terms of eq. (\ref{dpsi}) and thus prevent that the developments of advanced quantum information processors reveals them.

A second delicate point about eq. (\ref{dpsi}) is that it is not relativistic and it seems impossible to make it relativistic \cite{noNLvariable}. 

The previous two points are part of the reasons I left the field some 20 years ago.

\section{conclusion}
I want to understand Nature. For me this requires that ``I'' exist and that there is something out there to be understood, in particular that there is a world out there. Physics is all about extracting information about How Nature Does It. For physicists “extracting information” means performing measurements. Hence the measurement problem has to be taken seriously. It is a real physics problem and its solution will provide new physics and new and testable predictions.

Taking standard Hilbert-space quantum theory at face value, without the vague collapse postulate, leads to the many-worlds: everything that can happen happens. The problem, besides hyper-determinism, is that ``I'' am excluded from the many-worlds. In order to re-introduce the ``I'', one has to introduce some effective collapses that happen when ``I'' interact with the world. Note that this step is usually not taken explicitly by the many-worlds followers, except when they compute predictions, i.e. when they do physics. This is a bit similar to the well-known Wigner friend story \cite{WignerFriend}, though Wigner never presented it in a many-worlds context. Hence, it seems that in order to make physical sense of many-worlds, one needs some form of dualism: ``I'' trigger effective collapses. Before me, everything co-existed. Now that I am here, in order to make predictions, I have to condition these ``co-existing things'', on those that correlated to me, using some effective collapses. Since it is a fact that ``I'' exist, wouldn't it be much simpler and cleaner to assume that the effective collapses are truly real and to include them in our physics theory?

Bohmian mechanics is a nice and constructive existence proof of non-local hidden variables. But it suffers from similar drawbacks than the many-worlds. It is hyper-deterministic and in order to make predictions one has to introduce an ``I'' that does some conditioning by de facto effective collapses. Hence, again, it is cleaner to assume real collapses in our physics theory. Moreover, doing so we may at least be wrong, i.e. at least we may predict new phenomena.

Remains the question of what triggers the collapses. Should we formulate the measurement problem as a search for those configurations of atoms and photons that trigger a collapse, as formulated in section \ref{QMP}? This is an interesting line of experimental research.

Dualism is a very natural position in our culture. Actually, I don't see how to avoid it for our Science to make sense. But I believe much premature to jump to the conclusion that it is the interaction between ``I'' and the outside world that triggers the collapses of the quantum states. Other forms of dualism, actually trialism: ``I'' plus two sorts of stuff out there, are logical possibilities, but there are no good candidates and introducing some new stuff seems too high a price to pay, especially when it is not (yet?) needed.

Remains spontaneous collapses, described for instance by some modified Sch\"odinger equation to which one adds some non-linear stochastic terms, as in eq. (\ref{dpsi}). These additional terms lead continuously and spontaneously, i.e. by mere random chance, to collapses that barely affect microscopic systems, but quickly localize macroscopic objects. It seems to me that the scientific method that has been that efficient so far tells us that this spontaneous collapse approach is by far the most promising one. For me, it is also the only one that is consistent with what I expect from physics. Indeed, at the end of the day, a theory without collapses, doesn't predict any events, hence has zero explanatory power.

One additional value of collapse theories is that they naturally incorporate the passage of time. I am well aware that it is fashion in physics to claim that time is an illusion \cite{Barbour99}. Admittedly, time is a complex notion, or series of notions with many facets, time may be relative, difficult to grasp, etc. But time exists. Moreover, time passes \cite{NortonTimePasses, SmolinTimePasses, GisinTimePasses}. 

With spontaneous collapse theories, time exists and passes, the world out there exists and undergoes a stochastic evolution. And ``I'' exist, outside the theory, able to contemplate it, to develop it and act as an observer.

\small
\section*{Acknowledgment} This work profited from stimulating discussions with Florian Fr\"owis and Renato Renner. Financial support by the European ERC-AG MEC and the Swiss NSF are gratefully acknowledged.

\end{document}